\documentstyle[12pt]{article}
\newcommand{\nc}{\newcommand}

\nc{\dbar}{\bar{\partial}}
\nc{\be}{\begin{equation}}
\nc{\ee}{\end{equation}}

\catcode`\@=11

\@addtoreset{equation}{section}

\def\@normalsize{\@setsize\normalsize{15pt}\xiipt\@xiipt
\abovedisplayskip 14pt plus3pt minus3pt%
\belowdisplayskip \abovedisplayskip
\abovedisplayshortskip  \z@ plus3pt%
\belowdisplayshortskip  7pt plus3.5pt minus0pt}
\def\small{\@setsize\small{13.6pt}\xipt\@xipt
\abovedisplayskip 13pt plus3pt minus3pt%
\belowdisplayskip \abovedisplayskip
\abovedisplayshortskip  \z@ plus3pt%
\belowdisplayshortskip  7pt plus3.5pt minus0pt
\def\@listi{\parsep 4.5pt plus 2pt minus 1pt
            \itemsep \parsep
            \topsep 9pt plus 3pt minus 3pt}}

\def\underline#1{\relax\ifmmode\@@underline#1\else
        $\@@underline{\hbox{#1}}$\relax\fi}
\@twosidetrue
\relax

\catcode`@=12

\catcode`\@=11

\def\section{\@startsection{section}{1}{\z@}{3.5ex plus 1ex minus
   .2ex}{2.3ex plus .2ex}{\large\bf}}

\def\ps@headings{\def\@oddfoot{}\def\@evenfoot{}
\def\@oddhead{\hbox{}\hfill
        \makebox[.5\textwidth]{\raggedright\ignorespaces --\thepage{}--
        \hfill }}
\def\@evenhead{\@oddhead}
\def\subsectionmark##1{\markboth{##1}{}}
}

\ps@headings

\catcode`\@=12

\relax

\def\figcap{\section*{Figure Captions\markboth
        {FIGURECAPTIONS}{FIGURECAPTIONS}}\list
        {Fig. \arabic{enumi}:\hfill}{\settowidth\labelwidth{Fig. 999:}
        \leftmargin\labelwidth
        \advance\leftmargin\labelsep\usecounter{enumi}}}
 \relax
\def\tablecap{\section*{Table Captions\markboth
        {TABLECAPTIONS}{TABLECAPTIONS}}\list
        {Table \arabic{enumi}:\hfill}{\settowidth\labelwidth{Table 999:}
        \leftmargin\labelwidth
        \advance\leftmargin\labelsep\usecounter{enumi}}}
 \relax
\def\reflist{\section*{References\markboth
        {REFLIST}{REFLIST}}\list
        {[\arabic{enumi}]\hfill}{\settowidth\labelwidth{[999]}
        \leftmargin\labelwidth
        \advance\leftmargin\labelsep\usecounter{enumi}}}
 \relax

\catcode`\@=11

\def\ps@headings{\def\@oddfoot{}\def\@evenfoot{}
\def\@oddhead{\hbox{}\hfill
        \makebox[.5\textwidth]{\raggedright\ignorespaces --\thepage{}--
        \hfill }}
\def\@evenhead{\@oddhead}
\def\subsectionmark##1{\markboth{##1}{}}
}

\ps@headings

\relax

\def\firstpage#1#2#3#4#5#6{
\begin{document}

\begin{titlepage}
\nopagebreak
\title{\begin{flushright}
       \vspace*{-1.8in}
       {\normalsize SISSA-159/96/EP} 
\end{flushright}
\vfill
{\large \bf #3}}
\author{\large #4 \\ #5}
\maketitle
\vskip -7mm
\nopagebreak
\begin{abstract}
{\noindent #6}
\end{abstract}
\vfill
\begin{flushleft}
\rule{16.1cm}{0.2mm}\\[-3mm]
\end{flushleft}
\footnotesize{PACS: 11.25.Mj} \\
\footnotesize{Keywords: F-terms, Strings, N=2, Duality}
\thispagestyle{empty}
\end{titlepage}}
\newcommand{\dal}{\raisebox{0.085cm}
{\fbox{\rule{0cm}{0.07cm}\,}}}
\newcommand{\dt}{\partial_{\langle T\rangle}}
\newcommand{\dtbar}{\partial_{\langle\bar{T}\rangle}}
\newcommand{\al}{\alpha^{\prime}}
\newcommand{\mst}{M_{\scriptscriptstyle \!S}}
\newcommand{\mpl}{M_{\scriptscriptstyle \!P}}
\newcommand{\dv}{\int{\rm d}^4x\sqrt{g}}
\newcommand{\lv}{\left\langle}
\newcommand{\rv}{\right\rangle}
\newcommand{\ph}{\varphi}
\newcommand{\sbar}{\,\bar{\! S}}
\newcommand{\xbar}{\,\bar{\! X}}
\newcommand{\fbar}{\,\bar{\! F}}
\newcommand{\zbar}{\,\bar{\! Z}}
\newcommand{\tbar}{\bar{T}}
\newcommand{\ubar}{\bar{U}}
\newcommand{\ybar}{\bar{Y}}
\newcommand{\phb}{\bar{\varphi}}
\newcommand{\cm}{Commun.\ Math.\ Phys.~}
\newcommand{\pr}{Phys.\ Rev.\ D~}
\newcommand{\prl}{Phys.\ Rev.\ Lett.~}
\newcommand{\pl}{Phys.\ Lett.\ B~}
\newcommand{\ibar}{\bar{\imath}}
\newcommand{\jbar}{\bar{\jmath}}
\newcommand{\np}{Nucl.\ Phys.\ B~}
\newcommand{\e}{{\rm e}}
\newcommand{\gsi}{\,\raisebox{-0.13cm}{$\stackrel{\textstyle
>}{\textstyle\sim}$}\,}
\newcommand{\lsi}{\,\raisebox{-0.13cm}{$\stackrel{\textstyle
<}{\textstyle\sim}$}\,}
\date{}
\firstpage{95/XX}{3122}
{\large\sc $N=2$ Type I-Heterotic Duality and Higher Derivative F-Terms} 
{Marco Serone}
{\normalsize\sl International School for Advanced Studies, ISAS-SISSA
\\[-3mm]
\normalsize\sl Via Beirut n. 2-4, 34013 Trieste, Italy\\[-3mm]
\normalsize\sl and\\[-3mm]
\normalsize\sl Istituto Nazionale di Fisica Nucleare, sez.\ di Trieste,
Italy\\[-3mm]
\normalsize\sl e-mail:serone@sissa.it \\[-3mm]}
{We test the conjectured Type I-Heterotic Duality in four dimensions by
analyzing a given class of higher derivative F-terms of the form
$F_gW^{2g}$, with $W$ the $N=2$ gravitational superfield.
We study a particular dual pair of theories, the $O(2,2)$ heterotic 
model and a type I model based on the $K3$ ${\bf Z_2}$ orbifold 
theory constructed
by Gimon and Polchinski, further compactified on a torus.
The $F_g$ couplings appear at 1-loop on both theories;
because of the weak-weak nature of this duality in four dimensions,
it is meaningful to compare the heterotic $F_g$'s with the corresponding 
type I couplings perturbatively.
We compute the $F_g$'s in type I, showing that they receive contributions
only from $N=2$ BPS states and that in the appropriate limit
they coincide with the heterotic couplings, in agreement with the given 
duality.} 

\section{Introduction}
One of the most important results achieved in the last two years, towards
an understanding of non-perturbative string theory, has been the 
equivalence of string theories previously considered as different. 
One of these equivalences
has been proposed by Polchinski and Witten in \cite{pw}, where they gave 
some arguments leading to the conclusion that type I and heterotic theory 
are dual to each other. Although in ten dimensions this is a strong-weak 
duality, it turns out that after suitable compactifications leading to 
supersymmetric $N=2$ theories in four dimensions, there exist regions in the
moduli space of the theory in which both the heterotic and the type I
descriptions are simultaneously weakly coupled.

In four dimensions $N=2$ string theories present another duality,
realized by dual pairs constructed from Calabi-Yau and $K3\times T^2$
compactifications of the type II and heterotic string respectively \cite{kv}.
While there have been several and strong checks for the aforementioned 
type II-heterotic duality \cite{agnt,klkk}, only recently 
the study of the four dimensional consequences of type I-heterotic 
duality has been started in ref.\cite{abf}. They studied the 1-loop 
corrections to the prepotential for a class of models obtained by 
compactifying type I on $T^4/{\bf Z_2}\times T^2$ \cite{gp}.
In particular it has been shown in \cite{abf} that the 
perturbative prepotential for a rank four model that admits type I, type II
and heterotic descriptions, agree.

The aim of this note is to compute in type I, for the rank four model,
1-loop amplitudes arising from the higher derivative F-terms of the form 
$F_g(X)W^{2g}$ where $W$ is the $N=2$ gravitational superfield and $X$
denotes generically the chiral vector superfields
of the theory. They are then compared, in the appropriate limit, to the 
corresponding 
heterotic couplings that have been already computed in \cite{agnt} for
a $O(2,1)$ model (and whose result has been  extended
in \cite{ms} for more general $O(2,n)$ models).

This note is organized as follows. In the next section we briefly
review the construction of the type I-heterotic dual pair under 
consideration, looking for the appropriate limit in the vector moduli
space in which the two
1-loop amplitudes have to coincide.
In section three we recall the form of the $F_g$ couplings 
in the heterotic theory and then,
in section four, we compute these terms in the type I side, showing that 
they are in complete agreement with the heterotic one's, providing a 
further check for the given duality. 

\section{The dual pair}

In this section we shortly remind how to construct a specific 
type I-heterotic rank four dual pair. On the heterotic side it is obtained
by compactifying the $E_8\times E_8$ theory on $K3\times T^2$, putting
12 instantons on each $E_8$, and then higgsing completely the remaining
two $E_7$'s. This leaves us with a massless spectrum containing  244 
hypermultiplets and just 
the three vectormultiplets coming from the torus \cite{kv}. Note that since
one vector-field is the graviphoton, the vector moduli space is parametrized
by three complex scalars, usually denoted S-T-U, where
\be S_H=\theta+ie^{-2\phi_H}, \ \ \ T_H=B_{45}+i\sqrt{G}, \ \ \
U_H=(G_{45}+i\sqrt{G})/G_{44} \ee
denote the complex dilaton, the K\"ahler and complex structure of the torus,
respectively \footnote{From now on a subscript H or I refers to heterotic 
and type I quantities.}. As already mentioned in the introduction, the 
type I dual is constructed starting from the six-dimensional models on a
$K3$ orbifold $T^4/{\bf Z_2}$ studied in \cite{gp}. The closure of the 
operator product expansion and the cancellation of tadpole divergencies
require the presence of 32 9-branes and 32 5-branes. Because of U(1) 
anomalies in the theory, putting two 5-branes at each fixed point and giving
non-vanishing vacuum expectation values to scalars,
it is possible to break completely the gauge group, remaining only with 244
$N=1$ hypermultiplets in six dimensions \cite{abf,blp}, 4 of which coming 
from the closed string sector and the remaining 240 from the 
open string sector.
After the further compactification on $T^2$, the massless spectrum consists
of 4 U(1)'s, 3 complex scalars and the 244 $N=2$ hypermultiplets, of course.
It is convenient in type I to take the following combinations of the three
complex scalars:
\be S_I=\theta+ie^{-\phi_I}G^{1/4}\omega^2, \ \ \
S^{'}_I=B_{45}+ie^{-\phi_I}G^{1/4}\omega^{-2}, 
\ \ \ U_H=(G_{45}+i\sqrt{G})/G_{44} \ee
where $\omega^4$ is the volume of the $K3$ orbifold. It is now 
straightforward to show that starting from the ten-dimensional relations 
$\phi^{10}_I=-\phi^{10}_H$ and 
$\alpha^{\prime}_I=e^{\phi^{10}_H}\alpha^{\prime}_H$
\cite{pw}, the duality map becomes \cite{abf}
\be  S_H=S_I, \ \ \ T_H=S_I^{'}, \ \ \ U_H=U_I \label{map} \ee 
So weakly coupled type I theory corresponds to weakly coupled heterotic 
theory provided that the volume of the $T^2$ torus is large.
It is important to note that the type I low-energy effective action is 
actually invariant under the two Peccei-Quinn symmetries 
$Re\,S_I\rightarrow Re\,S_I$ + const., 
$Re\,S_I^{'}\rightarrow Re\,S_I^{'}$ + const., implying the perturbative 
independence of chiral amplitudes (in loops) from $S_I$ and $S_I^{'}$.
On the other hand the heterotic effective action presents only the usual
P-Q symmetry, $Re\,S_H\rightarrow Re\,S_H$ + const. Given the relations 
(\ref{map}), this immediately implies that for every chiral amplitude $A$:
\be \lim_{T_2\rightarrow\infty}A_H(T,U)=A_I(U)|_{S_2>S_2^{\prime}} 
\label{lim} \ee
where $T_2=Im\,T$ \footnote{The rescriction $S_2>S_2^{\prime}$ is due to 
the fact that in the heterotic theory the large $S_H$ limit is taken before
the $T_H$ limit \cite{abf}.}.

\section{Heterotic $F_g$ couplings in the 
$T_2\rightarrow\infty$ limit}
We briefly review in this section the computation done in \cite{agnt} of
these couplings, in order to take the 
$T_2\rightarrow\infty$ limit. The relevant amplitude for $F_{g,H}$
involves two gravitons and $2g-2$ graviphotons. 
It has been shown in \cite{agnt} that these amplitudes contain only
correlators of space-time bosons that can be easily computed
by defining a generating function for the $F_{g,H}$'s 
\footnote{This is slightly different from the generating function defined
in \cite{agnt}, in order to compare better the heterotic results with the 
type I one's.},
$F_H(\lambda,T,U)\equiv\sum_{g=1}^{\infty}g^2\lambda^{2g}F_{g,H}(T,U)$. 
This allows the exponentiation of the operators, 
reducing in this way the computation to the evaluation of a determinant.
From the results given by \cite{agnt}, it is not difficult to show that
\be F_H(\lambda,T,U)=\frac{\lambda^2}{\pi^2}\int\frac{d^2\tau}{\tau_2}
\frac{1}{\bar{\eta}^4(\bar{q})}C(\bar{q})
\sum_{n_1, n_2\atop m_1, m_2}
q^{\frac{1}{2}|P_L|^2}\bar{q}^{\frac{1}{2}|P_R|^2}
\frac{1}{2}\frac{d^2}{d\tilde{\lambda}^2}\left[\left(
\frac {2\pi i\tilde{\lambda}\bar{\eta}^3(\bar{q})}
      {\bar{\theta}_1(\tilde{\lambda},\bar{\tau})}    \right)^2
e^{-\frac{\pi\tilde{\lambda}^2}{\tau_2}}             \right]
\label{fht} \ee
where $q=e^{2i\pi\tau}$
\begin{eqnarray} 
P_L&=&\frac{1}{\sqrt{2T_2U_2}}(n_1+n_2\bar{U}+m_1\bar{T}+m_2\bar{T}\bar{U})
\nonumber \\
P_R&=&\frac{1}{\sqrt{2T_2U_2}}(n_1+n_2\bar{U}+m_1T+m_2T\bar{U})
\end{eqnarray}
$\tilde{\lambda}=P_L\tau_2\lambda/\sqrt{2T_2U_2}$, 
$\bar{\theta}_1$ is the odd theta-function and $C(\bar{q})$ is
the partition function of $K3$ in the odd spin structure, that can be
shown to depend only on $\bar{q}$.
In particular, $C(\bar{q})$ has the following expansion \cite{agnt}:
\be C(\bar{q})=\bar{q}^{-5/6}(1-244\,\bar{q}+...) \ee
where the first factor accounts for the tachyon and 244 is the number of 
hypermultiplets of the model.
In order to show more explicitly the $T_2$ dependence of 
$F_H(\lambda)$ rescale $\tau_2\rightarrow T_2\tau_2/2$; then, bringing
the limit inside the integral and expanding in $\bar{q}$, the non-vanishing
result will be given by the $\bar{q}^0$ coefficient of the expansion: 
\be
F_H(\lambda,U)=\lim_{T_2\rightarrow\infty}F_H(\lambda,T,U)=
\frac{\lambda^2}{2\pi^2}\int_0^{\infty}\frac{d\tau_2}{\tau_2}
\sum_{n_1,n_2}e^{-\frac{\pi\tau_2|P|^2}{2U_2}} 
\left[240\frac{d^2}{d\lambda^2} \left(\frac{\lambda\pi}
{\sin\bar{\lambda}} \right)^2+16\pi^2\right] \ee
where $P\equiv n_1+n_2\bar{U}$ and $\bar{\lambda}=\lambda\pi\tau_2P/4U_2$.
The second term in square brackets comes from the tachyon contribution
together with the linear term in $\bar{q}$ deriving from the expansion
of the eta and theta-function and gives a contribution only to 
$F_1$\footnote{Note that $F_1$ should be evaluated by a three point function
\cite{agn} but, as showed by \cite{agnt}, the computation we have summarized
here gives the 
right result even for this case. The same will happen in Type I theory.}. 
We will see in the next section how this amplitude is reproduced in type I. 

\section{Computation of $F_g$ in the type I model}
We consider here the same amplitude involving two gravitons and $2g-2$ 
graviphotons; the graviton vertex operator is the usual
\be V^{\mu\nu}_g(p)=(\partial X^{\mu}+ip\cdot\psi\psi^{\mu})
(\bar{\partial}X^{\nu}+ip\cdot\psi\psi^{\nu})e^{ip\cdot X} \ee
while the graviphoton vertex operator is obtained applying the two
supersymmetric charges to $V_g$:
\be V_{\gamma}(p)=(Q_1^{(L)}+Q_1^{(R)})(Q_2^{(L)}+Q_2^{(R)})V_g(p)\ee
where
\begin{eqnarray}
Q_{\alpha,1}^{(L)}+Q_{\alpha,1}^{(R)}&=&\oint dz
e^{-\frac{\phi}{2}}
S_{\alpha}\Sigma e^{i\frac{H_5}{2}}(z)+\oint d\bar{z}
e^{-\frac{\tilde{\phi}}{2}}\tilde{S}_{\alpha}\tilde{\Sigma}
e^{i\frac{\tilde{H}_5}{2}}(\bar{z}) \nonumber \\
Q_{\alpha,2}^{(L)}+Q_{\alpha,2}^{(R)}&=&\oint dz
e^{-\frac{\phi}{2}}
S_{\alpha}\bar{\Sigma}e^{i\frac{H_5}{2}}(z)+\oint d\bar{z}
e^{-\frac{\tilde{\phi}}{2}}\tilde{S}_{\alpha}\tilde{\bar{\Sigma}}
e^{i\frac{\tilde{H}_5}{2}}(\bar{z}) \end{eqnarray}
and $e^{-\frac{\phi}{2}}$, $e^{i\frac{H_5}{2}}$ are the bosonization
of the superghosts and of the complex fermion associated to the internal 
torus respectively, $S_{\alpha}$ is the space-time spin field operator and 
$\Sigma$ and its complex conjugate are the $K3$ internal spin field
operators; bosonizing the U(1) Cartan current in the SU(2) algebra 
of the internal $N=(4,4)$ SCFT as $J_3=i\sqrt{2}H$, $\Sigma$ can be 
written as $\Sigma=e^{i\frac{\sqrt{2}}{2}H}$. The same of course 
for the right-moving sector. The 1-loop amplitude involves a sum over the 
torus, Klein bottle, annulus and M\"obius strip surfaces. 
We may now use the results of ref.\cite{bm}, in order to compute 
the boson and fermion propagators on all the surfaces starting from
those on the torus,
by the method of images; similarly to what found in \cite{gjn}
(and following the same notation), it is then 
possible to see that after having extracted the right dependence 
on momenta and summed over the spin structures over 
all the surfaces, the $F_g$ couplings
are given by the following amplitude in the odd spin structure:
\begin{eqnarray}
&\!&F_{g,I}(U)=
\frac{1}{(g!)^2\pi^2}\sum_{\alpha=\rm T,\rm K\atop\rm M,\rm A}
\int [dM]_{\alpha}\,C_\alpha([t])\sum_{n_1, n_2}
e^{-\frac{\pi [t]|P|^2}{U_2\sqrt{G}}}(\frac{P}{2U_2\sqrt{G}})^{2g-2}
\langle V_g^+V_g^-\prod_{i=1}^{g-1}\prod_{j=1}^{g-1} \nonumber \\
\hspace{-1cm}&\!&\left[\int\! d^2x_i(Z_1^+(\partial+
\bar{\partial})Z_2^+\!+\!(\psi_1^+-\tilde{\psi}_1^+)
(\psi_2^+-\tilde{\psi}_2^+)\right]
\left[\int\! d^2y_j(Z_2^-(\partial+
\bar{\partial})Z_1^-\!+\!(\psi_1^--\tilde{\psi}_1^-)
(\psi_2^--\tilde{\psi}_2^-)\right]\rangle_\alpha \nonumber
\end{eqnarray}
where \be 
V_g^{\pm}=\int\! d^2x(Z_{1,2}^{\pm}\partial Z_{2,1}^{\pm}+
\psi_1^{\pm}\psi_2^{\pm})(Z_{1,2}^{\pm}\bar{\partial}
Z_{2,1}^{\pm}+\tilde{\psi}_1^{\pm}\tilde{\psi}_2^{\pm}) \ee
$[dM]_\alpha$ denotes the measure of the moduli integration for each 
surface $\alpha$ with $[t]$ the corresponding coordinate, $C_\alpha([t])$
is the partition function in the odd spin structure of the internal sector
and $P=n_1+n_2\bar{U}$ are the discrete momenta corresponding to the 
$T^2$ torus. 
Define now \be F_I(\lambda,U)\equiv\sum_{g=1}^{\infty}g^2\lambda^{2g} 
F_{g,I}(U)\ee Exponentiating we obtain
\be\hspace{-1cm}
F_I(\lambda,U)=\frac{\lambda^2}{\pi^2}\sum_{\alpha=\rm T,\rm K\atop\rm 
M,\rm A} \int [dM]_{\alpha}\,C_\alpha([t])\sum_{n_1, n_2}
e^{-\frac{\pi t|P|^2}{U_2\sqrt{G}}}\langle e^{-S_0+\tilde{\lambda}S}
V_g^+V_g^-\rangle_{\alpha} \ee
where $\tilde{\lambda}=\lambda tP/\sqrt{2U_2}G^{1/4}$, $S_0$ is the free 
action for the space-time bosons and fermions and
\be S=\int\frac{d^2x}{t}[
Z_1^+(\partial+\bar{\partial})Z_2^++(\psi_1^+-\tilde{\psi}_1^+)
(\psi_2^+-\tilde{\psi}_2^+)+
Z_2^-(\partial+\bar{\partial})Z_1^-+(\psi_1^--\tilde{\psi}_1^-)
(\psi_2^--\tilde{\psi}_2^-)] \ee
Because of the four zero modes $\psi_1^{\pm}=\tilde{\psi}_1^{\pm}=$const.,
$\psi_2^{\pm}=\tilde{\psi}_2^{\pm}=$const. of the new action, it is easy
to check that
\be \langle e^{-S_0+\tilde{\lambda}S}
V_g^+V_g^-\rangle_{\alpha}=\frac{t^2}{2}\frac{d^2}{d\tilde{\lambda}^2}         
\langle e^{-S_0+\tilde{\lambda}S}\rangle_{\alpha} \ee
The amplitude is then reduced to the evaluation of determinants
of space-time bosons and fermions; before computing them, however,
note that $C([t])$ is actually an index on all the surfaces.
In order to see this, it is better to go to the operatorial formalism.
For the torus and annulus, $C({t})$ coincides with the Witten index
\cite{witt} ${\rm Tr}_{\rm R-R}(-)^{F_L+F_R}q^{L_0}\bar{q}^{\bar{L}_0}$ and 
${\rm Tr}_{\rm R}(-)^{F}q^{L_0}$. For the M\"obius strip and Klein bottle, 
$C_{\alpha}(t)$ is respectively ${\rm Tr}_{\rm R}\Omega (-)^{F}q^{L_0}$ and
${\rm Tr}_{\rm R-R}\Omega q^{L_0}\bar{q}^{\bar{L}_0}$. 
It is easy to see that these are still indices. In the M\"obius strip
$\Omega$ will act simply by multiplying each multiplet by a common
eigenvalue, while in the Klein bottle it is possible to check that each
multiplet entering in the evaluation of the trace has equal number
of states with opposite eigenvalues of $\Omega$. 
Since the $C_{\alpha}([t])$ are all indices, their values are invariant
for small perturbations of the theory and then it is possible to compute 
them directly from the spectrum of the free theory considered in \cite{gp}.
By a simple counting in the closed string spectrum, we easily derive that
$C_{\rm K}=0$ and $C_{\rm T}=8$ \footnote{Note that the 
twisted closed states in the orbifold limit are all massive \cite{blp}.}. 
Since in the torus the fermion and boson determinants cancel, 
leaving a $\lambda$-independent constant, we have a non-vanishing
contribution only for the coupling $F_{1,I}$ by the following term:
\be F_{1,I}^{\rm torus}=\frac{1}{\pi^2}\int d^2\tau\, C_{\rm T} \sum_{n_1, 
n_2} e^{-\frac{\pi \tau_2|P|^2}{U_2\sqrt{G}}}
\langle V_g^+V_g^-\rangle_{\rm T} \ee
The only non-vanishing contribution comes when we take all the eight 
fermions of the two graviton vertex operators to soak up the eight 
zero-modes present in the odd spin structure of the torus. 
The result gives simply:
\be F_{1,I}^{\rm torus}=4\int \frac{d^2\tau}{\tau_2}\, C_{\rm T} 
\sum_{n_1, n_2} e^{-\frac{\pi \tau_2|P|^2}{U_2\sqrt{G}}} \ee
Rescaling $\tau_2\rightarrow \sqrt{G}\,\tau_2/2$,
we obtain, for large $\sqrt{G}$:
\be F_{1,I}^{\rm torus}=32\int_0^{\infty} \frac{d\tau_2}{\tau_2} 
\sum_{n_1, n_2} e^{-\frac{\pi \tau_2|P|^2}{2U_2}} \ee
The remaining contribution for the $F_g$'s comes from the annulus and 
M\"obius strip. The corresponding determinants give equal contribution,
with the following result: 
\be \langle e^{-S_0+\tilde{\lambda}S}\rangle_{\rm M}=
\langle e^{-S_0+\tilde{\lambda}S}\rangle_{\rm A} =
\frac{1}{t^3}\prod_{m=1}^{\infty}\left(1-\frac{\tilde{\lambda}^2}
{m^2}\right)^{-2}=\frac{1}{t^3}\left(\frac{\tilde{\lambda}\pi}
{\sin\tilde{\lambda}\pi}\right)^2 \ee
Looking to the open string spectrum of \cite{gp}, it turns out
that $C_{\rm A}+C_{\rm M}=32^2-2\cdot 32=4\cdot 240$.
Putting everything together and rescaling $t\rightarrow \sqrt{G}\,t/2$, we
finally obtain:
\be
F_I(\lambda,U)=\frac{2\lambda^2}{\pi^2}\int_0^{\infty}\frac{dt}{t} 
\sum_{n_1,n_2}e^{-\frac{\pi t|P|^2}{2U_2}}\left[240
\frac{d^2}{d\lambda^2}\left(\frac{\lambda\pi}{\sin\bar{\lambda}}
\right)^2+16\pi^2\right] \ee
where $\bar{\lambda}=\lambda\pi tP/4U_2$, reproducing, up to a constant,
the generating function $F_H(\lambda,U)$, according to the type 
I-heterotic duality.
Note also that only $N=2$ BPS states give contributions to the $F_g$ 
couplings, like in the heterotic case \cite{hm}.

{\bf{Acknowledgements}}

It is a pleasure to thank J.F. Morales, E. Gava, K.S. Narain and
M.H. Sarmadi for useful and enlightening discussions.


\newpage
\begin{thebibliography}{99}
\bibitem{pw} J. Polchinski and E. Witten, {\it Evidence for 
Heterotic-Type I String Duality}, Nucl. Phys. {\bf B460} (1996) 525;
\bibitem{kv} S.Kachru and C.Vafa, {\it Exact Results for N=2 
Compactifications of Heterotic Strings}, Nucl. Phys. {\bf B450} (1995) 69;
\bibitem{agnt} I. Antoniadis, E. Gava, K.S. Narain and T.R. Taylor,
{\it N=2 Type II-Heterotic Duality and Higher Derivative F-terms},
Nucl. Phys {\bf B455} (1995) 109;
\bibitem{klkk} V. Kaplunovsky, J. Louis and S. Theisen, {\it Aspects
of Duality in N=2 String Vacua}, Phys. Lett. {\bf B357} (1995) 71;
S. Kachru, A. Klemm, W. Lerche, P. Mayr and C. Vafa, 
{\it Non-perturbative results on the point particle limit of N=2
Heterotic String Compactifications}, Nucl. Phys. {\bf B459} (1996) 537; 
\bibitem{abf} I. Antoniadis, C. Bachas, C. Fabre, H. Partouche and T.R. 
Taylor, {\it Aspects of Type I-Type II-Heterotic Triality in Four
Dimensions}, hep-th/9608012;
\bibitem{gp} E.G. Gimon and J. Polchinski, {\it Consistency Conditions
for Orientifolds and D-Manifolds}, Phys. Rev. {\bf D54} (1996) 1667;
\bibitem{ms} J.F. Morales and M. Serone, {\it Higher Derivative F-terms 
in N=2 Strings}, Nucl. Phys. {\bf B481} (1996) 389;
\bibitem{blp} M. Berkooz, R.G. Leigh, J. Polchinski, J.H. Schwarz,
N. Seiberg and E. Witten, {\it Anomalies, Dualities and Topology of 
D=6 N=1 Superstrings Vacua}, hep-th/9605184;
\bibitem{agn} I. Antoniadis, E. Gava and K.S. Narain, {\it Moduli
Corrections to Gravitational Couplings from String Loops}, Phys. Lett.
{\bf B283} (1992) 209; {\it Moduli Corrections to Gauge and Gravitational 
Couplings in four-dimensional Superstrings}, Nucl. Phys. {\bf B383} 
(1992) 109; 
\bibitem{bm} C.P. Burgess and T.R. Morris, {\it Open and Unoriented Strings
\`a la Polyakov}, Nucl. Phys. {\bf B291} (1987) 256; {\it Open 
Superstrings \`a la Polyakov}, Nucl. Phys. {\bf B291} (1987) 285;
\bibitem{gjn} E. Gava, T.Jayaraman, K.S. Narain and M.H. Sarmadi, 
{\it D-Branes and the Conifold Singularity}, hep-th/9607131;
\bibitem{witt} E. Witten, {\it Constraints on Supersymmetry Breaking},
Nucl. Phys. {\bf B202} (1982) 253;
\bibitem{hm} J.A. Harvey and G. Moore, {\it Algebras, BPS States and 
Strings}, Nucl. Phys. {\bf B463} (1996) 315.
\end{thebibliography}
\end{document}